\documentclass[12pt,aps,prb,tightenlines,showpacs]{revtex4}
\usepackage{times}
\usepackage[T1]{fontenc}
\usepackage[latin1]{inputenc}
\usepackage{amsmath}
\usepackage{amsfonts}
\usepackage{latexsym}
\usepackage{graphicx}
\usepackage{amssymb}
\usepackage{subfigure}
\usepackage[flushleft]{caption2}

\makeatletter

\providecommand{\LyX}{L\kern-.1667em\lower.25em\hbox{Y}\kern-.125emX\@}

\usepackage{amsfonts}

\makeatother

\begin{document}


\title{Quantum generalized constant coupling model for geometrically frustrated antiferromagnets}

\author{A. J. García-Adeva}

\email{garcia@landau.physics.wisc.edu}

\author{D. L. Huber}

\affiliation{Department of Physics; University of Wisconsin--Madison; Madison, WI 53706}

\begin{abstract}
A generalized constant coupling approximation for quantum geometrically frustrated
antiferromagnets is presented. Starting from a frustrated unit, we introduce
the interactions with the surrounding units in terms of an internal effective
field which is fixed by a self consistency condition. Results for the static
magnetic susceptibility and specific heat are compared with previous results
in the framework of this same model for the classical limit. The range of applicability
of the model is discussed.
\end{abstract}
\maketitle

\section{Introduction}

\begin{figure}
\centering
\subfigure[\label{fig.pyro.struc}Pyrochlore lattice.]
{\includegraphics{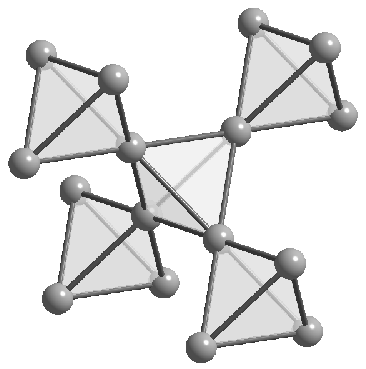}}
\hspace{1in}
\subfigure[\label{fig.kagome.struc}\emph{Kagom\'{e}} lattice.]
{\includegraphics{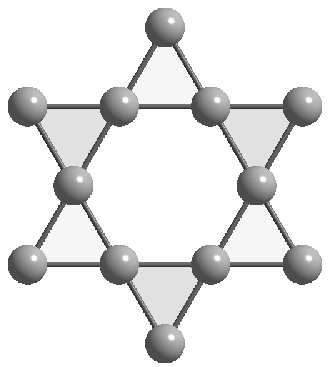}}
\caption{\label{fig.structures}The magnetic lattices considered in this work.}
\end{figure}

In the last several years, geometrically frustrated antiferromagnets (GFAF)
have emerged as a new class of magnetic materials with uncommon physical properties,
and have received a great deal of attention (see Refs.~\onlinecite{hfm2000,Ramirez1994,Schiffer1996a}
and references therein). In these materials, the elementary unit of the magnetic
structure is the triangle, which makes it impossible to satisfy all the antiferromagnetic
bonds at the same time, with the result of a macroscopically degenerate ground
state. Examples of GFAF are the pyrochlore and the \emph{kagomé} lattices. In
the former, the magnetic ions occupy the corners of a 3D arrangement of corner
sharing tetrahedra; in the later, the magnetic ions occupy the corners of a
2D arrangement of corner sharing triangles (see Fig.~\ref{fig.structures}).
In the case of materials which crystallize in the pyrochlore structure, the
static magnetic susceptibility follows the Curie--Weiss law down to temperatures
well below the Curie--Weiss temperature. At this point, usually one to two orders
of magnitude smaller than the Neél point predicted by the standard mean field
(MF) theory, some systems exhibit some kind of long range order (LRO), whereas
others show a transition to a spin glass state (SG). This is a striking feature
for a system with only a marginal amount of disorder. Finally, there are some
pyrochlores which do not exhibit any form of order whatsoever, and are usually
regarded as spin liquids. In the case of the \emph{kagomé} lattice, even though
there are very few real systems where this structure is realized, the magnetic
properties fall in two major categories: the vast majority of the compounds
studied show a transition to a LRO state with a non collinear configuration
of spins, and a few systems exhibit no LRO, but a SG like transition.

For these reasons, it is easy to understand the large amount of attention, both
from the experimental and the theoretical points of view, these systems have
attracted. From the theoretical point of view, a number of techniques have been
used to try to understand the origin of the puzzling properties mentioned above.
All the theoretical results seem to indicate that the classical Heisenberg model
with only nearest neighbor interactions does not display any long range order
for these geometries, in agreement with Monte Carlo (MC) results\cite{Moessner99, Moessner98a, Moessner98b, Moessner98,Reimers1991, Harris92, Reimers1992, Chalker92a,Reimers1993,Garanin99}. 

There are also a relatively few works which have dealt with the quantum effects
in these systems\cite{Harris91, Canals98, Canals00a, Isoda98, CHUBUKOV1992, GARCIA-ADEVA2000a, ELSTNER94, ZENG1990, NAKAMURA95},
even though the main interest during last years has been on the classical GFAF.
However, in a recent work\cite{GARCIA-ADEVA2000a}, the present authors showed
that there are some features of the experimental data for the susceptibility
in pyrochlore compounds that can only be understood in a quantum framework,
as, for example, the maxima appearing in this quantity at temperatures well
below the Curie--Weiss temperature.

In another recent work\cite{GARCIA-ADEVA2000b}, the present authors developed
a generalization of the well known constant coupling method\cite{KASTELEJEIN1956}
(CC), which can be applied to frustrated geometries, the so called generalized
constant coupling (GCC) method. This technique was applied successfully to the
study of the classical Heisenberg Hamiltonian with nearest neighbors (NN) interactions
in both the pyrochlore and \emph{kagomé} lattices. In spite of the mathematical
simplicity of this technique, the results obtained for the susceptibility are
essentially exact when compared with Monte Carlo data in both types of lattices.
Moreover, the calculated specific heat is also in very good agreement with Monte
Carlo data, even though there are some deviations at very small temperatures,
which can be understood in the light of the fact that a MF theory cannot properly
describe this property (or, equivalently, the internal energy) at \( T=0 \)
K, due to the distinct nature of the excitations which are important in this
limit, namely, the spin waves. In any case, the GCC method provides an excellent
starting point for studying the magnetic properties of the frustrated systems
in the paramagnetic region.

The first question that arises is which features of these magnetic properties
are different in a quantum formulation of the GCC method, and that is precisely
what we try to answer in this work. However, in contrast with the classical
limit, there are no reliable quantum Monte Carlo calculations for these systems,
due to well known difficulties that arise in the quantum version of this technique,
which makes it very difficult to check the accuracy of the present quantum version
of the GCC method. It is tempting to compare the results of this method with
experimental data available in the literature. However, here we are considering
the most restrictive approach to the problem by focusing on Heisenberg models
with NN interactions only, whereas in real systems, as pointed out above, there
are always additional effects, so we think such a comparison makes no sense
at this point. Therefore, the most we can do is to give the results for the
quantum version of the GCC method, and study how they evolve towards the classical
results for large values of the individual spin quantum number. If this transition
from the quantum to the classical limit is smooth, we can expect the predictions
of the model to be, at least, reasonable in the paramagnetic region.

\section{The model}

The Heisenberg model with only nearest neighbor (NN) interactions in the presence
of a magnetic applied field \( H_{0} \) is described by the Hamiltonian\cite{SMART1966}
\begin{equation}
H=J\sum _{\left\langle i,j\right\rangle }\vec{s}_{i}\cdot \vec{s}_{j}-H_{0}\sum _{i}s_{z_{i}},
\end{equation}
 where \( J \) is the positive antiferromagnetic coupling, \( \vec{s}_{i} \)
and \( \vec{s}_{j} \) represent quantum spins of modulus \( s_{0} \) located
in a pyrochlore or \emph{kagomé} lattice and \( s_{z_{i}} \) the corresponding
component along the applied field, and the sum is done over pairs of NN.

The idea of our approximate method is based on the experimental fact that the
spin-spin correlations in the GFAF lattices are short ranged\cite{Moessner99}.
Therefore, it is a reasonable approximation to start by considering isolated
units (tetrahedra or triangles, for the pyrochlore and \emph{kagomé} lattices,
respectively), and later add the interactions with the surrounding units in
an approximate way. Thus, it is important to first study the properties of the
individual units. This task has been carried out by García--Adeva and Huber
for the quantum case\cite{GARCIA-ADEVA2000a}, and we will not repeat the derivation
here. It will suffice to say that the susceptibility per spin of an isolated
unit with \( p \) spins (\( p=4 \) for the pyrochlore and \( p=3 \) for the
\emph{kagomé}) is given by\begin{equation}
\chi _{p}(T)=\frac{\left\langle S_{p}^{2}\right\rangle }{3pT}=\frac{1}{3pT}\frac{\sum _{S}g(S)S(S+1)(2S+1)e^{-jS(S+1)/2}}{\sum _{S}g(S)(2S+1)e^{-jS(S+1)/2}},
\end{equation}
 where \( j=J/T \), and \( \left\langle S_{p}^{2}\right\rangle  \) represents
the average value of the square of the total spin of the unit, and \( g(S) \)
is the number of configurations with total spin \( S \) (see Table \ref{table.degeneracies}).
\begin{table}
{\centering \begin{tabular}{cc|ccccccccccc}
\hline 
\hline 
 &
&
\multicolumn{11}{|c}{\( S \) }\\
&
&
\( \frac{1}{2} \)&
\( \frac{3}{2} \)&
\( \frac{5}{2} \)&
\( \frac{7}{2} \)&
\( \frac{9}{2} \)&
\( \frac{11}{2} \)&
\( \frac{13}{2} \)&
\( \frac{15}{2} \)&
&
&
\\
&
&
~\( 0 \)~&
~\( 1 \)~&
~\( 2 \)~&
~\( 3 \)~&
~\( 4 \)~&
~\( 5 \)~&
~\( 6 \)~&
~\( 7 \)~&
~\( 8 \)~&
~\( 9 \)~&
~\( 10 \)~\\
\hline 
\( p=3 \)&
\( s_0=\frac{1}{2} \)&
\( 2 \)&
\( 1 \)&
&
&
&
&
&
&
&
&
\\
&
\( s_0=1 \)&
\( 1 \)&
\( 3 \)&
\( 2 \)&
\( 1 \)&
&
&
&
&
&
&
\\
&
\( s_0=\frac{3}{2} \)&
\( 2 \)&
\( 4 \)&
\( 3 \)&
\( 2 \)&
\( 1 \)&
&
&
&
&
&
\\
&
\( s_0=\frac{5}{2} \)&
\( 2 \)&
\( 4 \)&
\( 6 \)&
\( 5 \)&
\( 4 \)&
\( 3 \)&
\( 2 \)&
\( 1 \)&
&
&
\\
\hline 
\( p=4 \)&
\( s_0=\frac{1}{2} \)&
\( 2 \)&
\( 3 \)&
\( 1 \)&
&
&
&
&
&
&
&
\\
&
\( s_0=1 \)&
\( 3 \)&
\( 6 \)&
\( 6 \)&
\( 3 \)&
\( 1 \)&
&
&
&
&
&
\\
&
\( s_0=\frac{3}{2} \)&
\( 4 \)&
\( 9 \)&
\( 11 \)&
\( 10 \)&
\( 6 \)&
\( 3 \)&
\( 1 \)&
&
&
&
\\
&
\( s_0=\frac{5}{2} \)&
\( 6 \)&
\( 15 \)&
\( 21 \)&
\( 24 \)&
\( 24 \)&
\( 21 \)&
\( 15 \)&
\( 10 \)&
\( 6 \)&
\( 3 \)&
\( 1 \)\\
\hline 
\hline 
\end{tabular}\par}

\caption{\label{table.degeneracies}Values of \protect\( g(S)\protect \) for some representative
values of \protect\( s_{0}\protect \). Integer values of \protect\( S\protect \)
correspond to the pyrochlore case (\protect\( p=4\protect \)) for any value
of \protect\( s_{0}\protect \) and the \emph{kagomé} lattice for integer values
of \protect\( s_{0}\protect \). Half integer values of \protect\( S\protect \)
correspond to the \emph{kagomé} lattice (\protect\( p=3\protect \)) for half
integer values of \protect\( s_{0}\protect \).}
\end{table}

For some derivations, it is better to express all the thermodynamic quantities
in terms of the function \begin{equation}
\varepsilon _{p}(T)=\frac{\left\langle S_{p}^{2}\right\rangle }{ps_{0}(s_{0}+1)}-1.
\end{equation}
 Therefore, the susceptibility of the isolated unit can be expressed as\begin{equation}
\chi _{p}(T)=\frac{s_{0}(s_{0}+1)\left( 1+\varepsilon _{p}(T)\right) }{3T}.
\end{equation}
 It is important to notice that the only difference from the derivation carried
out in Ref.~\onlinecite{GARCIA-ADEVA2000b} is in definition of \( \left\langle S_{p}^{2}\right\rangle  \).
The rest of the method is essentially the same.

Next, the interaction with neighboring units is introduced as an unknown internal
effective field, \( H_{1} \), created by the \( (p-1) \) NN ions outside the
unit. The CC approximation consists of taking this internal field as \( H_{1}=(p-1)H' \),
where \( H' \) is the average internal field acting on a spin due to each of
its NN. By comparison, in the case of the standard CW model, the internal effective
field is given by \( H_{1}=2(p-1)H' \), as the ions are considered separately,
and each has \( 2(p-1) \) NN in the corner sharing structures considered in
this work (see fig.~\ref{fig.structures}). The internal field is evaluated
by imposing the self consistent condition of equating the magnetization per
spin in the field with that of a unit in the field. In the paramagnetic region,
and small applied field, the susceptibility per spin is then given by\cite{GARCIA-ADEVA2000b}\begin{equation}
\chi ^{gcc}_{p}(T)=\frac{s_{0}(s_{0}+1)}{3T}\frac{1+\varepsilon _{p}(T)}{1-\varepsilon _{p}(T)}.
\end{equation}
The corresponding expressions for the internal energy\footnote{For comparison,
the ground state energy obtained from \eqref{internal.energy} for the \emph{kagomé}
lattice for $s_0=1/2$ (in units of $J$) is $-1/2$, whereas the corresponding
value for the pyrochlore lattice is $-3/4$ for the same value of $s_0$.}and
the specific heat per spin are given by\begin{equation}
\label{internal.energy}
u_{p}(T)=Js_{0}(s_{0}+1)\varepsilon _{p}(T)
\end{equation}
 and\begin{equation}
c_{p}(T)=Js_{0}(s_{0}+1)\frac{\partial }{\partial T}\varepsilon _{p}(T),
\end{equation}
 respectively.

\section{Results for the susceptibility and the specific heat}

In this section, we will present the main results of the model introduced in
the previous section. One is tempted to compare them with the available experimental
data. However, such a comparison makes no sense at this point because, as stated
in the introduction, there are several effects not included in this simplified
model such as dipole--dipole interactions, anisotropies, or further neighbor
interactions, to cite some examples. Therefore, we will present the results
of the present model and study how the quantum case evolves towards the classical
limit for relatively large values of the individual spins. If the quantum behavior
of the physical quantities of interest (namely, susceptibility and specific
heat) is not so different from the classical one for relatively large values
of \( s_{0} \), we can feel confident that the predictions are essentially
correct, at least in the temperature region where deviations from the classical
behavior are small, as we know that the GCC model gives an accurate description
of these quantities in whole the temperature range. It would be desirable also
to compare the predictions of this model with the ones obtained from more sophisticated
techniques (quantum Monte Carlo, high temperature series expansions, density
matrix renormalization group, exact diagonalization, and son on). However, that
comparison is not possible in general, due to the geometrical complexity of
GFAF which prevents one from applying any of these techniques to arbitrary values
of \( s_{0} \), and these kind of calculations have been only carried out for
the quantum \( s_{0}=1/2 \) which, as we will see below, is the case which
we expect to have the worst results.
\begin{figure}
\centering
\subfigure[Magnetic susceptibility.]
{\includegraphics[width=2.75in]{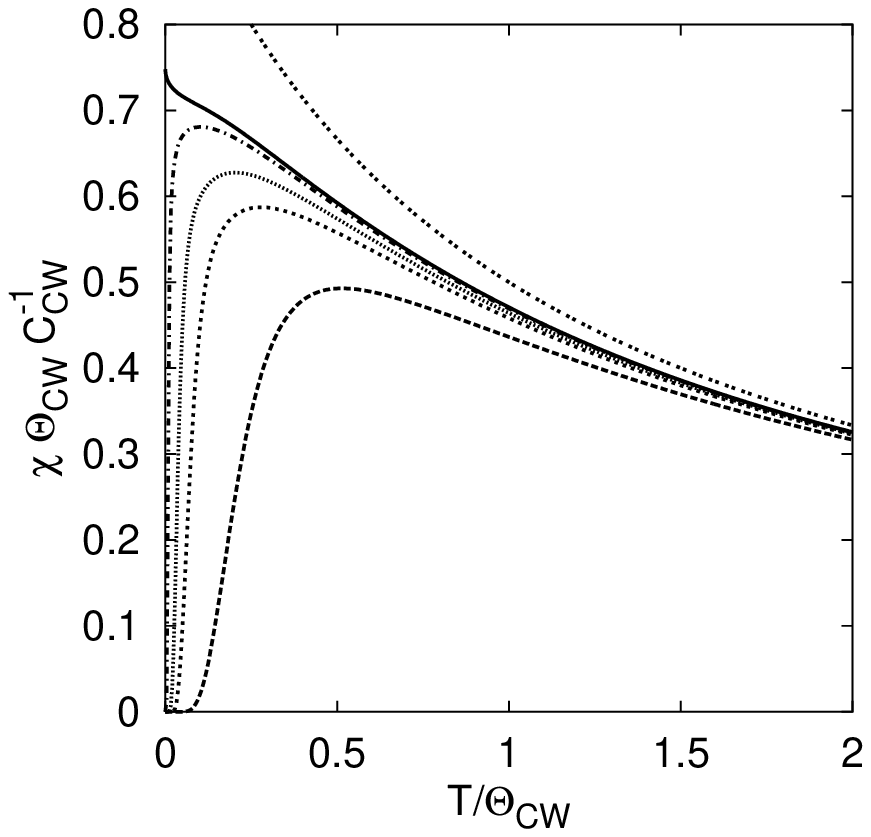}}
\subfigure[Specific heat.]
{\includegraphics[width=2.75in]{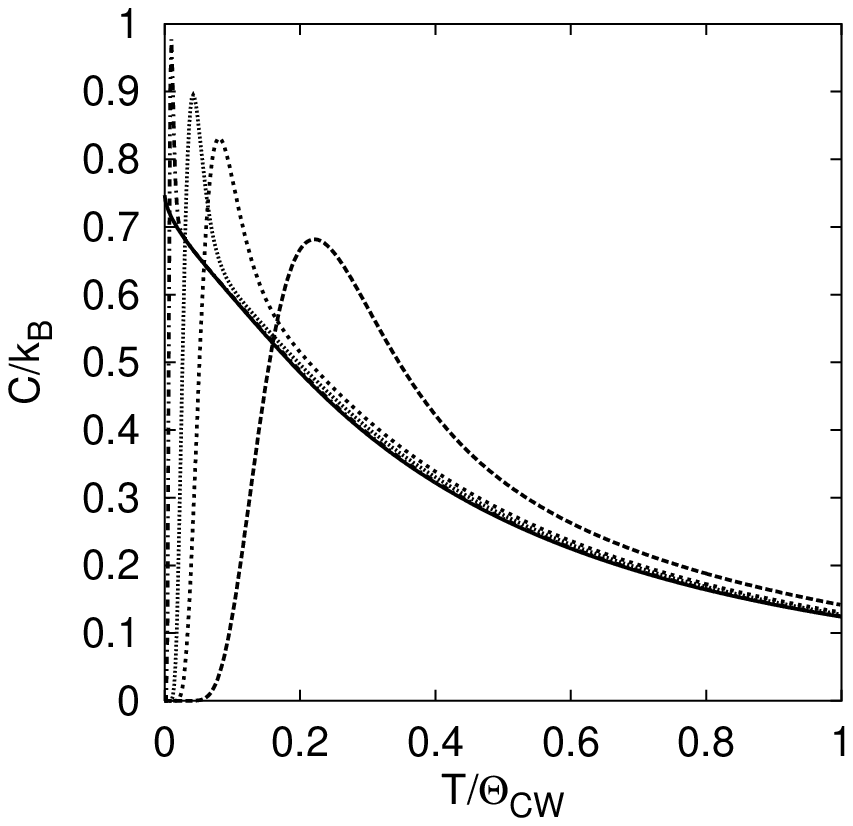}}
\caption{\label{fig.pyro}Susceptibility and specific heat for the pyrochlore lattice
for various values of \protect\( s_{0}\protect \): (long dashed line ) \protect\( s_{0}=1/2\protect \);
(short dashed line) \protect\( s_{0}=1\protect \); (dotted line) \protect\( s_{0}=3/2\protect \);
(dotted dashed line) \protect\( s_{0}=7/2\protect \) . The solid line is the
classical limit of the model. The double dotted line in the case of the susceptibility
represents the Curie--Weiss law.}
\end{figure}

\begin{figure}
\centering
\subfigure[Pyrochlore lattice. ($\bullet$) Maxima in the susceptibility; ($\circ$) Maxima in the specific heat.]
{\includegraphics[width=3in]{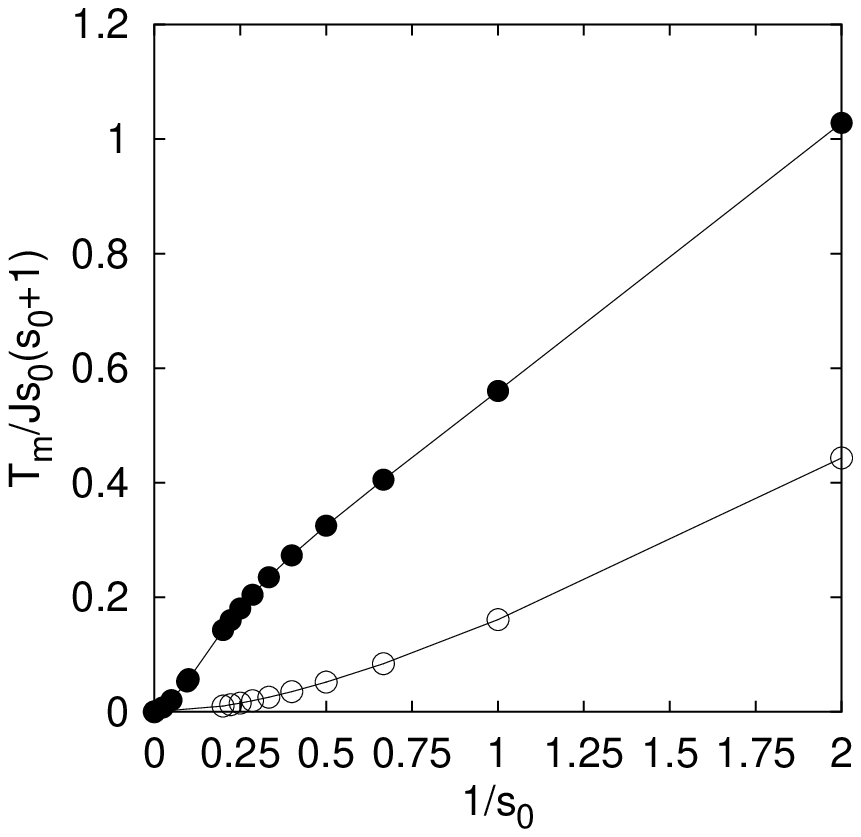}}
\subfigure[\emph{Kagom\'{e}} lattice. ($\Box$) Maxima in the susceptibility; ($\circ$) Maxima in the specific heat for half integer $s_0$; ($\times$) Maxima in the specific heat for integer values of $s_0$.]
{\includegraphics[width=3in]{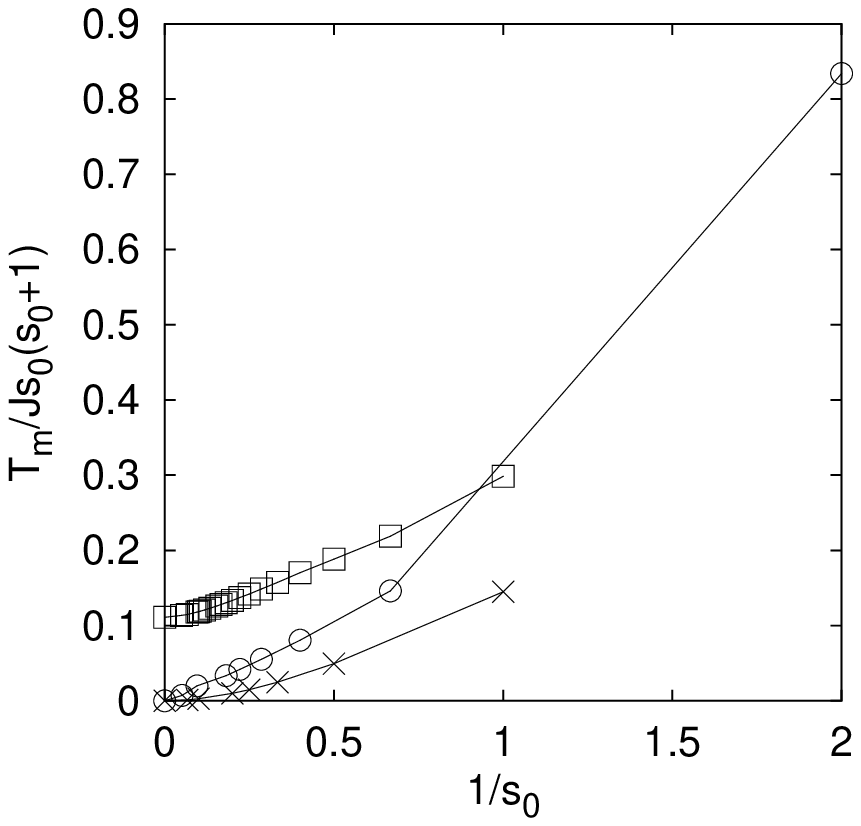}}
\caption{\label{fig.maxima}Position of the maxima appearing in the susceptibility and
specific heat. Solid lines are a guide to the eye.}
\end{figure}

Let us start by considering the results for the susceptibility and specific
heat in the pyrochlore lattice, which can be seen in Fig.~\ref{fig.pyro} for
some values of \( s_{0} \). As we can see from the observation of that figure,
the classical susceptibility deviates from the Curie--Weiss behavior below \( \Theta _{CW} \).
The quantum case is qualitatively similar to the classical behavior, except
at very low temperatures, where it passes through a maximum and then falls to
zero. However, we can also see that even for relatively small values of \( s_{0} \)
the quantum behavior is not so different from the classical one down to the
maximum. Regarding the predictions for the specific heat, again, the main difference
is that it goes through a maximum, and later falls to zero, in contrast with
the classical situation, where it goes to a constant value at \( T=0 \) K.
Additional information can be gained by studying the dependence of the position
of the maxima with the value of \( s_{0} \) in Fig.~\ref{fig.maxima}. In
that figure can be seen how nicely both the maxima in the susceptibility and
specific heat go to the zero classical value.

Let us now turn to the \emph{kagomé} lattice. In this case, it is reasonable
to expect the model to work less well than in the pyrochlore lattice, as it
is well known that the role of long wavelength fluctuations is more important
in 2D lattices, which are explicitly neglected in any MF approach. However,
from the MC results for the classical limit, it turns out that long wavelength
thermal fluctuations are unimportant for GFAF at finite temperatures, which
partially explains why the classical limit of our model is essentially exact
when compared to MC data for the susceptibility. The quantum case is a more
delicate one, as quantum fluctuations can play an important role. However, these
fluctuations will be only important at low temperatures, in a region where our
model does not apply for other reasons we will explain below.
\begin{figure}
\centering
\subfigure[Magnetic susceptibility.]
{\includegraphics[width=2.75in]{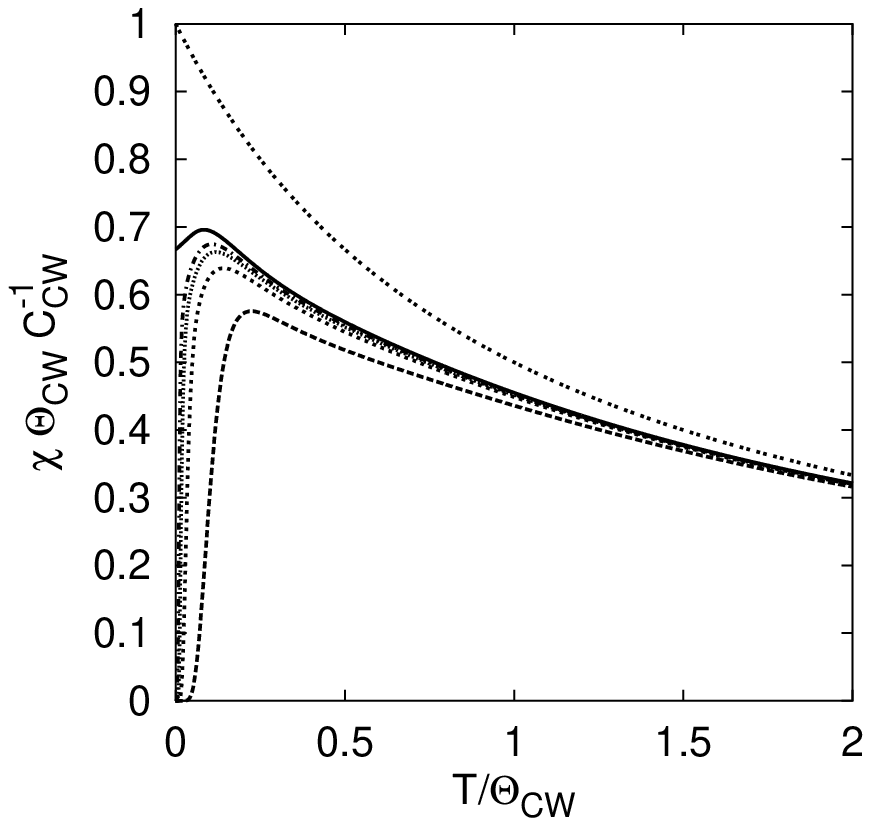}}
\subfigure[Specific heat.]
{\includegraphics[width=2.75in]{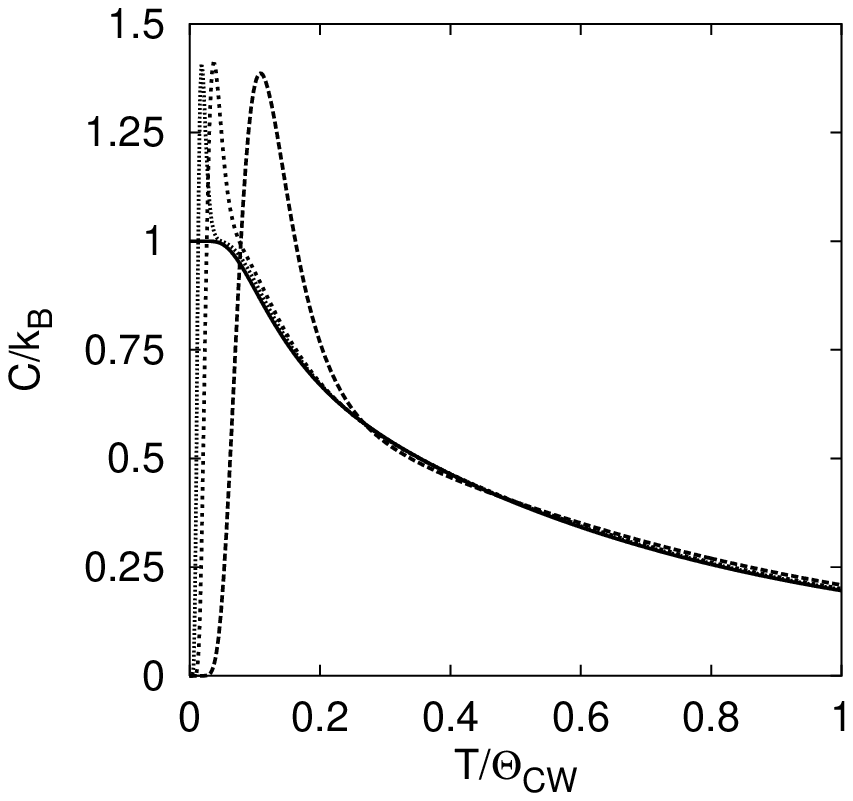}}
\caption{\label{fig.kagome.integer}Susceptibility and specific heat for the \emph{kagomé}
lattice with integer values of \protect\( s_{0}\protect \): (long dashed line)
\protect\( s_{0}=1\protect \); (short dashed line) \protect\( s_{0}=2\protect \);
(dotted line) \protect\( s_{0}=3\protect \); (dotted dashed line) \protect\( s_{0}=4\protect \).
The solid line is the classical limit of the model. The double dotted line in
the case of the susceptibility represents the Curie--Weiss law.}
\end{figure}

In order to proceed further, we will split the discussion for the \emph{kagomé}
lattice in two cases, corresponding to integer and half integer values of the
individual spins.

Let us first analyze the results for integer values of \( s_{0} \), which can
be seen in Fig.~\ref{fig.kagome.integer}. There are no remarkable differences
with respect to the discussion for the pyrochlore, except for the fact that
the maximum in the susceptibility is present even in the classical limit. The
only qualitative difference is the later fall to zero of both the susceptibility
and specific heat. Also, we have computed the evolution of the maxima for both
the susceptibility and specific heat. Again, both quantities go smoothly to
their classical values. It is important to stress that the fact that the susceptibility
goes to zero is related to the ground state total spin of the cluster, which,
for integer values of the individual spins, is always a singlet. This is not
the case for half integer values of \( s_{0} \).
\begin{figure}[h!]
\centering
\subfigure[Susceptibility per ion.]
{\includegraphics[width=3in]{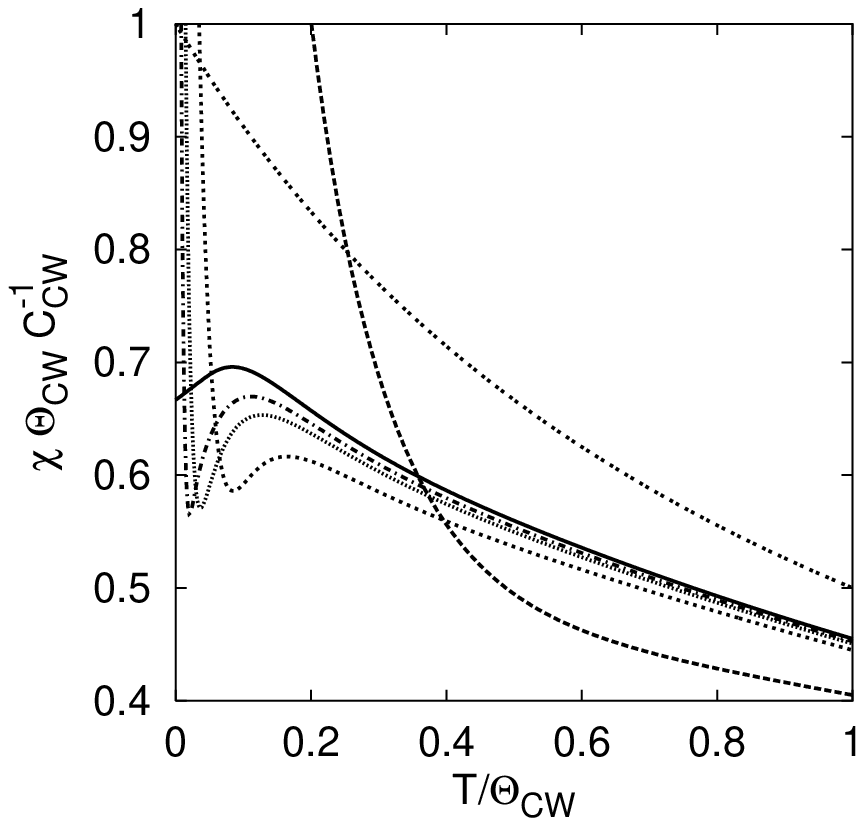}}
\subfigure[Specific heat per ion.]
{\includegraphics[width=3in]{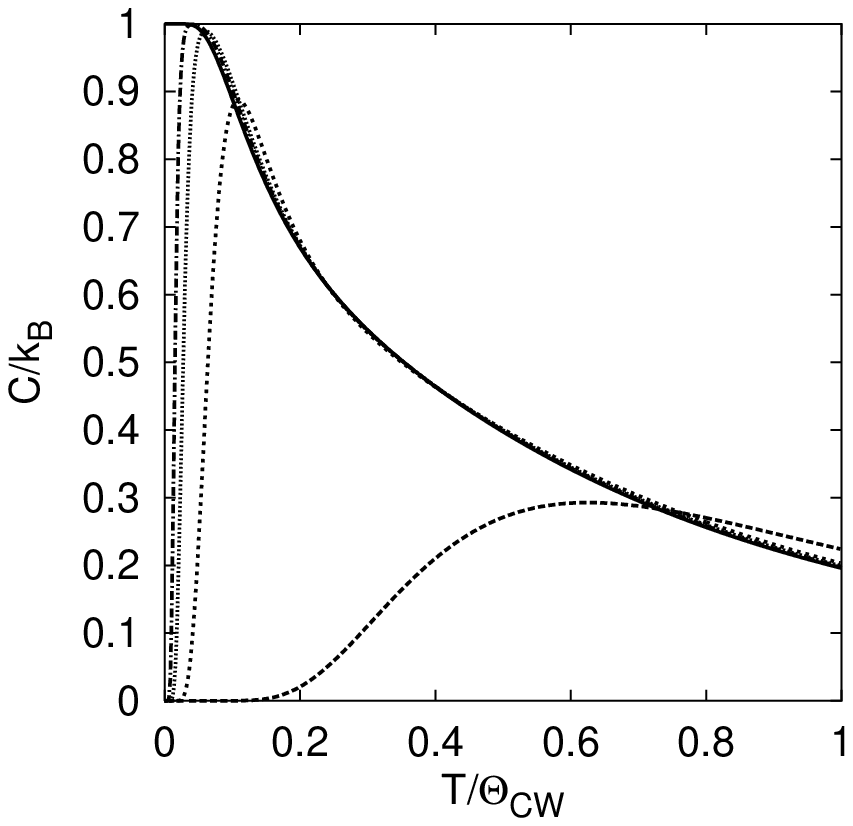}}
\caption{\label{fig.kagome.half.integer}Susceptibility and specific heat for the \emph{kagomé}
lattice with half integer values of \protect\( s_{0}\protect \): (long dashed
line) \protect\( s_{0}=1/2\protect \); (short dashed line) \protect\( s_{0}=3/2\protect \);
(dotted line) \protect\( s_{0}=5/2\protect \); (dotted dashed line) \protect\( s_{0}=7/2\protect \)
. The solid line is the classical limit of the model. The double dotted line
in the case of the susceptibility represents the Curie--Weiss law.}
\end{figure}

For half integer values of \( s_{0} \), there are very important differences
with respect to the previous cases, as is evident in Fig.~\ref{fig.kagome.half.integer}.
The susceptibility goes through a maximum, then reaches a minimum, and diverges
as \( T\to 0 \). This is a consequence of the fact that the ground state of
the triangular cluster for half integer values of \( s_{0} \) is a doublet
in our model. However, it is now generally accepted that the real ground state
of the \emph{kagomé} lattice is a singlet, and may have a gap in the spectrum
so, actually, the susceptibility should either be finite or go to zero as \( T\to 0 \).
Obviously, the present breakdown at very low temperatures is not a particular
feature of the mean field approach, but it is shared by every model which starts
from finite cluster calculations, with an odd number of spins in the elementary
cluster. Moreover, the maxima in the specific heat follow a different trend
than in the integer \( s_{0} \) case, even though, the classical limit is correct
(see Fig.~\ref{fig.maxima}). Even more striking is the fact that both the
susceptibility and specific heat for \( s_{0}=1/2 \) are qualitatively different
from all the previous presented results. The susceptibility does not show a
maximum, but diverges as \( 1/T \) as \( T \) approaches 0. The maximum in
the specific heat does not follow the extrapolated behavior for other half integer
values of \( s_{0} \). These kinds of problems have been already pointed out
in the literature for more rigorous methods than ours. A qualitative argument
that can help to understand these deviations is the fact that \( s_{0}=1/2 \)
is the {}``most quantum{}'' case, in which the physical quantities are more
sensitive to the discrete structure of the energy levels of the system. Taking
in to account that for half integer values of \( s_{0} \), the energy level
structure of our model is qualitatively incorrect, we should not be surprised
by this special behavior.
\begin{figure}[h!]
\centering
\subfigure[Susceptibility.]
{\includegraphics[width=3in]{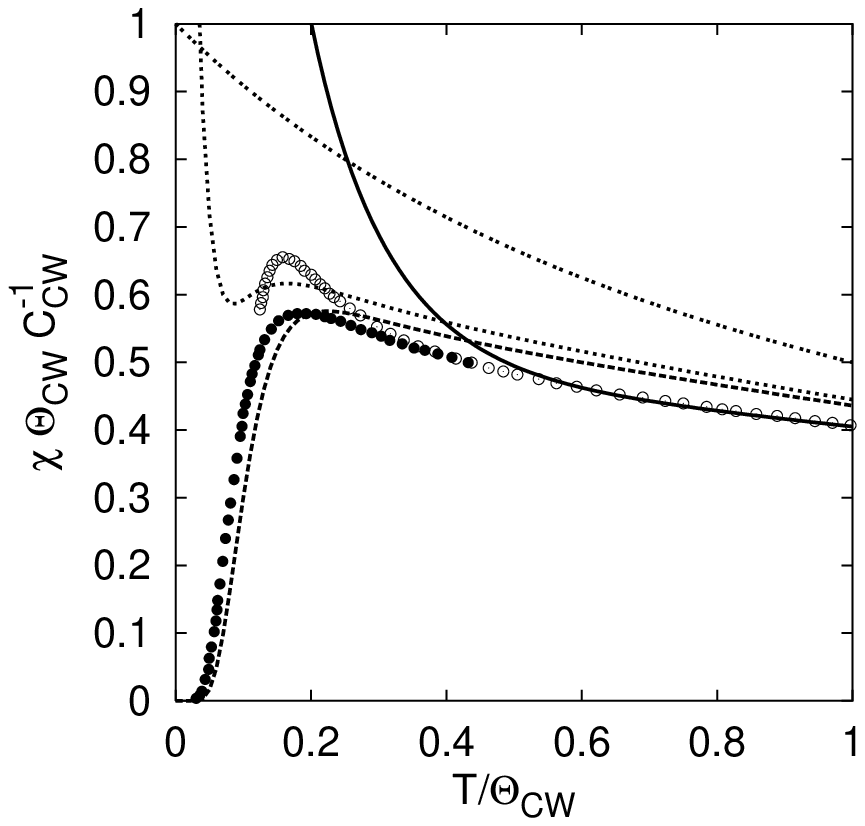}}
\subfigure[Specific heat.]
{\includegraphics[width=3in]{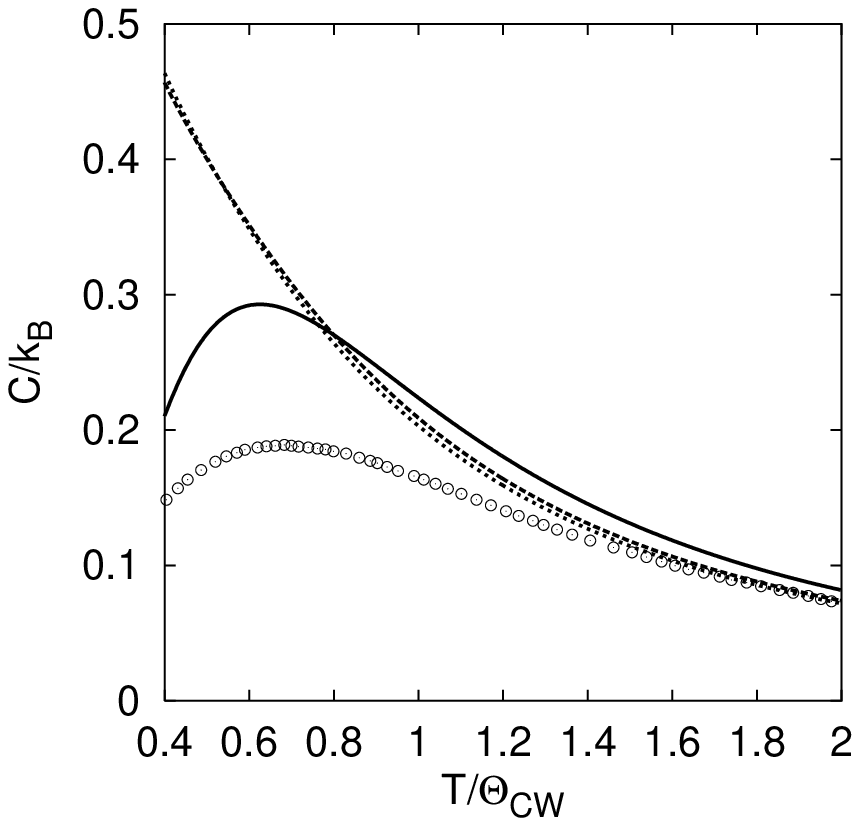}}
\caption{\label{fig.clusters}Comparison of the GCC predictions with results from finite
size clusters calculations and high temperature series for the \emph{kagomé}
1/2 lattice\cite{ELSTNER94}: (solid line) GCC model for \protect\( s_{0}=1/2\protect \);
(long dashed line) GCC model for \protect\( s_{0}=1\protect \); (dashed line)
GCC model for \protect\( s_{0}=3/2\protect \); ($\circ$) Results from high
temperature series expansions; ($\bullet$) Results from exact diagonalization
of a cluster with 18 spins. The double dotted line in the case of the susceptibility
represents the Curie--Weiss law. }
\end{figure}

As commented above, if we want to further check the range of applicability of
the model, it would be desirable to compare its predictions with the ones obtained
from more sophisticated methods, as exact diagonalization of small clusters
or high temperature series expansions. Unfortunately, these calculations have
been carried out only for the \emph{kagomé} with \( s_{0}=1/2 \)\cite{ELSTNER94, ZENG1990}
where, as we have argued above, we can expect the model to be least successful.
However, even though we do not expect complete agreement, we have carried out
such a comparison with results obtained from the aforementioned methods and
the results are presented in Fig.~\ref{fig.clusters}. In that figure we have
also plotted the corresponding curves corresponding to \( s_{0}=1 \) and \( s_{0}=3/2 \)
for comparison, at a qualitative level. It is important to stress that the curves
depicted there do not contain any fitting parameter nor have they been rescaled
in any sense. Surprisingly, the susceptibility calculated from our model for
\( s_{0}=1/2 \) is in very good agreement down to \( \approx 0.5\, \Theta _{CW} \)
with both high temperature series expansions and the diagonalization of a cluster
with 18 spins. The corresponding curve for the specific heat is in poor quantitative
agreement with the results from high temperature series expansions. However,
it describes accurately the position of the maximum. In any case, we can expect
the agreement with the specific heat to be worse, as it considers independent
units, neglecting the emerging correlations at low temperatures. Moreover, we
can see that the calculated susceptibilities for \( s_{0}=1 \) and \( s_{0}=3/2 \)
reproduce the main qualitative features of the high temperature expansion and
finite cluster results. At this point, it is difficult to say if this a fortunate
coincidence or an indication of the accuracy of our model for higher values
of \( s_{0} \). Numerical data for these values of the spins would be necessary
to decide between the two cases.

In any case, it is important to notice that some of the qualitative features
obtained in this simple model, have been observed in experimental studies. Again,
it is difficult to say at this point if those features are due to additional
interactions not accounted for in this model. For example, the existence of
a maximum in the susceptibility for the pyrochlore lattice has been systematically
observed in all the experiments at low enough temperatures. Actually, these
maxima where successfully interpreted by the present authors with an even simpler
model, in which the interactions with nearest neighbors and next nearest neighbors
were taken into account\cite{GARCIA-ADEVA2000a}. Moreover, in the work by Wills
and coworkers\cite{WILLS2000} on iron jarosites, which is one of the few systems
where the \emph{kagomé} lattice has an experimental realization, the magnetic
susceptibility exhibits the maximum and a later upturn, at a temperature comparable
with that predicted by our model. However, in that temperature region, it seems
likely that the dilution of the magnetic lattice by non magnetic impurities
plays an important role, giving rise to some kind of spin glass behavior, for
which it is very difficult to extract any conclusion.

In the light of these results, we think that a prudent estimate of the breakdown
of our model is set by the position of the maximum in the specific heat. As
we approach this temperature, emerging correlations that cannot be described
in the GCC formalism enter into play. For the pyrochlore lattice and the \emph{kagomé}
with integer values of \( s_{0} \), it is very likely that the present description
is qualitatively correct for temperatures well below the peak temperature of
the specific heat. However, we think that the description below that temperature
for half integer \( s_{0} \) in the \emph{kagomé} lattice is incorrect. Probably,
there is a maximum and a minimum in the susceptibility, but we expect the susceptibility
either vanishes or goes to a finite value at \( T\to 0 \), not to diverge.
In any case, it is still a significant improvement with respect to the Curie-Weiss
theory and other MF theories. However, more rigorous techniques are needed in
order to verify these assertions.

\section{Conclusions}

In this work we have presented the quantum version of the generalized constant
coupling model, which was shown by the authors to be in excellent agreement
with Monte Carlo data for the susceptibility and specific heat in the classical
limit, for both the pyrochlore and \emph{kagomé} lattices. There are some important
qualitative differences between the classical and quantum behaviors which are
important at low temperatures.

The main features of the susceptibility in this model are the presence of maxima
in both the susceptibility and specific heat, similar to those found experimentally.
The susceptibility for the \emph{kagomé} lattice with half integer values of
the individual spins is found to pass through a maximum and then, after a minimum,
diverges as \( T\to 0 \). This divergence is due to a non zero value of the
total spin of the ground state of the elementary units for these values of the
individual spins. There is some experimental evidence for the existence of such
an upturn in the susceptibility in the iron jarosite systems. However, the divergence
of this quantity in our model, which is due to the finite size cluster with
odd number of spins, is incorrect.

The results for the \emph{kagomé} \( 1/2 \) are compared with high temperature
series expansions and exact diagonalization of small clusters. The susceptibility
is found to be in good agreement with those results down to \( T\approx 0.5\, \Theta _{CW} \),
which is a remarkable fact for so simple a model. Even though the specific heat
is not in quantitative agreement at this temperature, the position of the maximum
is adequately predicted. Results from the aforementioned techniques for these
lattices for higher values of \( s_{0} \) would be desirable in order to check
the accuracy of the GCC model.

In any case, we feel that the present model provides an adequate description
of the nearest-neighbor \emph{kagomé} and pyrochlore systems down to the temperature
of the peak in the specific heat.

\end{document}